\documentclass[conference, a4paper]{IEEEtran}
\IEEEoverridecommandlockouts
\usepackage{cite}
\usepackage{amsmath,amssymb,amsfonts}
\usepackage{algorithmic}
\usepackage{algorithm}
\usepackage{graphicx}
\usepackage{textcomp}
\usepackage{xcolor}
\usepackage{cases}
\usepackage{amsmath}
\usepackage{graphicx}
\usepackage{subfigure}
\usepackage{geometry}
\newcommand{\mv}[1]{\mbox{\boldmath{$ #1 $}}}
\def\BibTeX{{\rm B\kern-.05em{\sc i\kern-.025em b}\kern-.08em
		T\kern-.1667em\lower.7ex\hbox{E}\kern-.125emX}}

\begin{document}
	
	\title{THz Beam Squint Mitigation via 3D Rotatable Antennas
	}
	
    \author{\IEEEauthorblockN{Yike Xie, Weidong Mei, Dong Wang, Boyu Ning, Zhi Chen, Jun Fang and Wei Guo}
	\IEEEauthorblockA{National Key Laboratory of Wireless Communications, \\
		University of Electronic Science and Technology of China (UESTC), Chengdu 611731, China
		\\
	Emails: ykxie@std.uestc.edu.cn; wmei@uestc.edu.cn; DongwangUESTC@outlook.com;\\  boydning@outlook.com; chenzhi@uestc.edu.cn; JunFang@uestc.edu.cn; guowei@uestc.edu.cn}
}

	\maketitle
	
	\begin{abstract}
	Analog beamforming holds great potential for future terahertz (THz) communications due to its ability to generate high-gain directional beams with low-cost phase shifters. However, conventional analog beamforming may suffer substantial performance degradation in wideband systems due to the beam-squint effects. Instead of relying on high-cost true time delayers, we propose in this paper an efficient three-dimensional (3D) rotatable antenna technology to mitigate the beam-squint effects, motivated by the fact that beam squint disappears along the boresight direction. In particular, we focus on a wideband wide-beam coverage problem in this paper, aiming to maximize the minimum beamforming gain within a given angle and frequency range by jointly optimizing the analog beamforming vector and the 3D rotation angles of the antenna array. However, this problem is non-convex and difficult to be optimally solved due to the coupling of the spatial and frequency domains and that of the antenna weights and rotation. To tackle this issue, we first reformulate the problem into an equivalent form by merging the spatial and frequency domains into a single composite domain. Next, we combine alternating optimization (AO) and successive convex approximation (SCA) algorithms to optimize the analog beamforming and rotation angles within this composite domain. Simulation results demonstrate that the proposed scheme can significantly outperform conventional schemes without antenna rotation, thus offering a cost-effective solution for wideband transmission over THz bands.
	\end{abstract}
	
	\section{Introduction}
	Terahertz (THz) communication, spanning a broad frequency range from $0.1$ THz to $10$ THz, has emerged as a promising technology to address the pressing spectrum congestion challenges in today's fifth-generation (5G) wireless communication systems\cite{chen1,ning1}. This ultra-wide bandwidth is expected to enable considerably higher data rates and lower latency compared to millimeter wave (mmWave) communications. However, THz signals suffer from a new beam-squint issue due to their large bandwidth\cite{ning3}. Specifically, in the hybrid precoding architecture, an analog beamformer can generate a directional beam aligned with the physical direction and achieve a full array gain. However, if the signal bandwidth keeps increasing, the beam will diverge from the desired physical direction at certain frequencies due to the frequency-independent phase shifters, which results in a significant loss in array gain. \par 
    To resolve the beam-squint issue, some existing works have proposed to deploy true time delayers (TTDs) to generate frequency-dependent phase shifts\cite{dai1}. However, due to the high cost of THz TTDs and the need to modify the existing hardware architecture of the phased array, their large-scale use faces difficulty in practice. Recently, fluid antennas (FAs) and movable antennas (MAs) have emerged as a new technology to improve wireless communication performance via local movement of antennas within a given region\cite{ma1,ma3,ma2,KKwong}. Compared with conventional fixed-position antennas (FPAs), MAs/FAs can leverage the new spatial degree of freedom to reshape wireless channels and alter the correlation among steering vectors in favor of wireless transmission. Inspired by their promising benefits, the performance of FA-/MA-aided systems has been extensively investigated under various system setups, e.g., secure communications\cite{mei1,cheng1}, multi-user systems\cite{zhu1}, cognitive radio\cite{weix1}, flexible beamforming\cite{beam_coverage,mamulti}, over-the-air computation (AirComp) network\cite{li1}. In \cite{6DMA}, the authors have proposed a more general six-dimensional movable antenna (6DMA) architecture for improving the wireless network capacity, by exploiting the three-dimensional (3D) antenna repositioning and 3D antenna rotation at the same time.\par
    	\begin{figure}[t]
		\centerline{\includegraphics[width=0.35\textwidth]{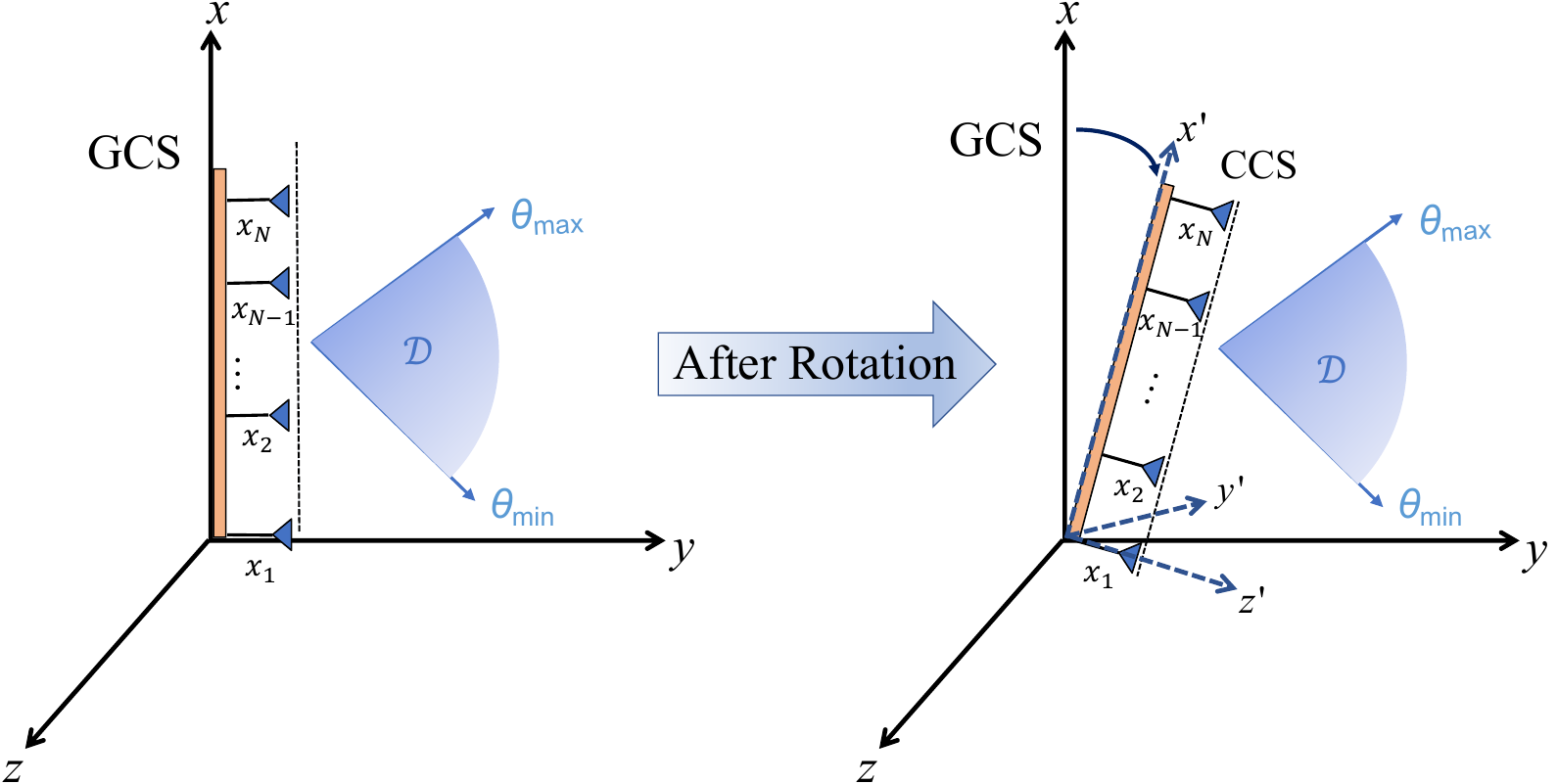}}
		\caption{3D rotatable antenna-assisted wideband wide-beam coverage.}\vspace{-12pt}
		\label{fig5}
		\end{figure}
	In this paper, we focus on leveraging the 3D rotation of a 6DMA array (thus termed rotatable antenna) to resolve the beam-squint issue over the THz band. This is mainly motivated by the phenomenon that the divergence of the beams over different frequencies becomes less/more severe as the user direction is close to/far away from the boresight of the antenna array. Hence, it is expected that by properly rotating the 3D angles of an antenna array, the beam-squint effects may be mitigated. To substantiate this claim, we investigate a wideband wide-beam coverage problem in this paper, aiming to maximize the minimum beamforming gain over a given angle and frequency range by jointly optimizing the analog beamforming vector and the 3D rotation angles of a linear rotatable antenna array, as shown in Fig. 1. However, the resulting problem is non-convex and generally difficult to be optimally solved due to the coupling of the spatial and frequency domains. To address this challenge, we develop an equivalent reformulation of the problem by merging the spatial and frequency domains into a single composite domain. Then, we optimize the analog beamforming and antenna rotation angles by combining AO algorithm and successive convex approximation (SCA). Numerical results show that our proposed scheme can achieve much better performance than conventional FPAs without rotation. It is worth noting that in a recent work\cite{zheng1}, the authors have conducted detailed performance analysis and optimization for rotatable antennas in a general multi-user communication system. Nonetheless, the investigation into rotatable antennas remains an early stage and its application to beam-squint mitigation has not been studied in the literature. It is also noted that compared to MAs/6DMAs, the rotatable antennas allow for a more efficient implementation in practice without the need for altering the relative positions of the antennas\cite{ning2}.\par 
	{\it Notations:} For a general matrix ${ \mv{W}}$, ${{ \mv{W}}^{ T}}$, ${{ \mv{W}}^{ H}}$ and ${{ \mv{W}}^\dag }$ denote its transpose, conjugate transpose, and conjugate, respectively. The symbols $\left|  \cdot   \right|$ and $\angle (  \cdot )$ denote the modulus and the angle of a complex number, respectively. The symbol $\left\|   \cdot  \right\|$ denotes the Euclidean norm of a vector. Moreover, $\text{diag}({ \mv{W}})$ denotes a diagonal matrix with the elements of ${ \mv{W}}$ on its diagonal. The sub-gradient of $f$ at $ \mv{A}$ is denoted by ${\partial _{ \mv{A}}}f$.
	
	\section{System Model and Problem Formulation}
	As shown in Fig. \ref{fig5}, we consider a wideband THz system, where a transmitter is equipped with a uniform linear array (ULA)\footnote{The proposed method can also be extended to the case of uniform planar array (UPA), as will be pursued in our follow-up work.} with $N$ antennas. Let $B$ and $f_c$ denote the total bandwidth and the carrier frequency of the system, respectively. In this paper, we focus on a wide-beam coverage problem, aiming to achieve a uniform beam gain over all directions within a given region in the angular domain (i.e., $\cal D$ in Fig. 1). Given the global coordinate system (GCS) in Fig. 1, we denote by $\theta_{\min}$ and $\theta_{\max}$ the boundary angles of $\cal D$, with $\theta_{\min}<\theta_{\max}$. To ease practical implementation, we consider analog beamforming at the transmitter. However, unlike the wide-beam coverage problem for the narrowband system with $B \ll f_c$, beam squint effects may occur due to the wideband transmission, leading to varying beam gains over the entire frequency band for any given direction in $\cal D$, as shown in Fig. 2(a). As a result, using conventional wide-beam designs for narrowband systems (see e.g., the references in \cite{ning1}) may incur significant performance loss. Nonetheless, it is known that more severe beam squint will appear in the direction with a large deviation from the boresight of the transmit array.
       \begin{figure}[t]
        \centering
        \subfigure[] {\includegraphics[width=0.20\textwidth]{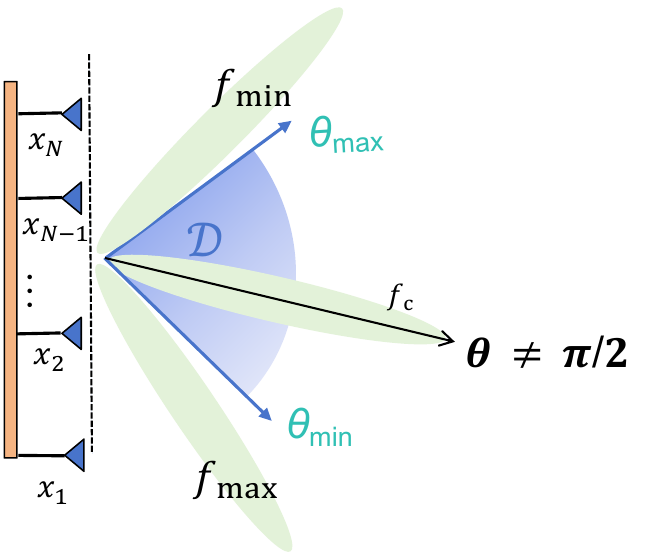}}
        \quad
        \subfigure[] {\includegraphics[width=0.20\textwidth]{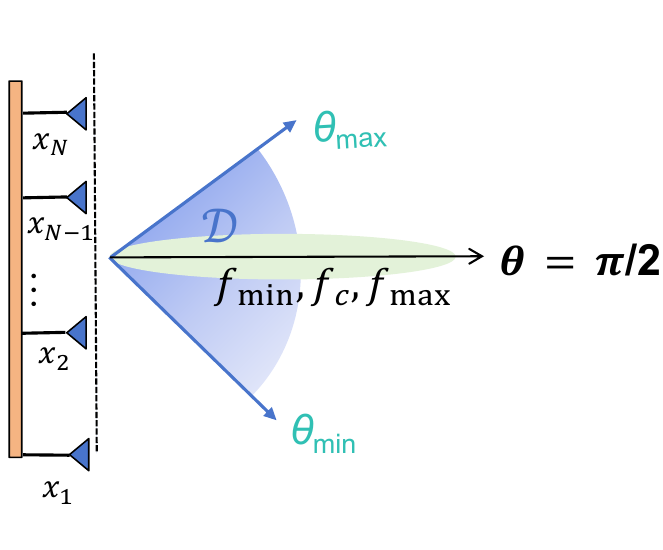}}
        \caption{Illustration of beam squint with different AoDs.}\label{simfig}\vspace{-16pt}
        
    \end{figure}
    In particular, when the target direction is just aligned with the boresight, i.e., the angle of departure (AoD) is equal to $\pi/2$, the beam squint will disappear, as shown in Fig. 2(b). Motivated by this fact and the recent advances in MAs and FAs, we propose in this paper a new approach to enable wideband wide-beam coverage by leveraging a 3D rotatable antenna array. To characterize the rotation of the antenna array, we establish a local coordinate system (LCS) for the ULA after its rotation and assume that it is parallel to the $x$-axis in the LCS, as shown in Fig.1. Then, the coordinate of its $n$-th antenna in the LCS can be expressed as
	\begin{equation}
		\mv{p}(n)=\left[x_n,0,0\right]^T,
	\end{equation}
	where $x_n=(n-1)d$ with $d$ representing the spacing between any two adjacent antenna elements. 
	Let $\mv{r}=[\alpha,\beta,\gamma]^T$ denote the rotational angle vector of the ULA from the GCS to the LCS, as shown in Fig. 1, where $\alpha \in [0,2\pi]$, $\beta \in [0,2\pi]$, $\gamma \in [0,2\pi]$ represent the rotational angles around the $x$-, $y$- and $z$-axes, respectively. \par 
	Notably, the relationship between the GCS and LCS can be characterized by three rotation matrices corresponding to the $x$-, $y$-, and $z$-axes, i.e.,
	\begin{equation}
		\mv{R}_x(\alpha ) = \left[ {\begin{array}{*{20}{c}}
				1&0&0\\
				0&{{c_\alpha }}&{ - {s_\alpha }}\\
				0&{{s_\alpha }}&{{c_\alpha }}
		\end{array}} \right],
	\end{equation}
	\begin{equation}
	\mv{R}_y(\beta ) = \left[ {\begin{array}{*{20}{c}}
			{{c_\beta }}&0&{{s_\beta }}\\
			0&1&0\\
			{ - {s_\beta }}&0&{{c_\beta }}
	\end{array}} \right],	
	\end{equation}
	\begin{equation}
		\mv{R}_z(\gamma ) = \left[ {\begin{array}{*{20}{c}}
				{{c_\gamma }}&{ - {s_\gamma }}&0\\
				{{s_\gamma }}&{{c_\gamma }}&0\\
				0&0&1
		\end{array}} \right],
	\end{equation}
	where we have defined ${c_\psi } = \cos (\psi )$ and ${s_\psi } = \sin (\psi )$ for notational simplicity, $\psi  \in \left\{ {\alpha ,\beta ,\gamma } \right\}$. Accordingly, the overall rotation matrix can be expressed as the product of the above three rotation matrices, i.e.,
	\begin{align}
		{ \mv{R}} &= {{ \mv{R}}_{x}}(\alpha )   {{ \mv{R}}_{y}}(\beta )   {{ \mv{R}}_{z}}(\gamma ) \\
	            	&= \left[ {\begin{array}{*{20}{c}}
				{{c_\alpha }{c_\gamma }}&{{c_\alpha }{s_\gamma }}&{ - {s_\alpha }}\\
				{{s_\beta }{s_\alpha }{c_\gamma } - {c_\beta }{s_\gamma }}&{{s_\beta }{s_\alpha }{s_\gamma } + {c_\beta }{c_\gamma }}&{{c_\alpha }{s_\beta }}\\
				{{{\mathop{\rm c}\nolimits} _\beta }{s_\alpha }{c_\gamma } + {s_\beta }{s_\gamma }}&{{{\mathop{\rm c}\nolimits} _\beta }{s_\alpha }{s_\gamma } - {s_\beta }{c_\gamma }}&{{c_\alpha }{c_\beta }}
		\end{array}} \right].
	\end{align}

	 Based on the above, the position of the $n$-th antenna of the ULA in the GCS can be determined as 
\begin{align}
	{\mv{{k}}(n)} =&{ \mv{R}}   {\mv{p}}(n)\\
	 =& \left[{{\mathop{\rm c}} _\alpha }{{\mathop{\rm c}} _\gamma }  ,{s_\alpha }{s_\beta }{c_\gamma } - {c_\beta }{s_\gamma }  ,{{\mathop{\rm c}} _\beta }{s_\alpha }{c_\gamma } + {s_\beta }{s_\alpha } \right]^T{x_n}.
\end{align}
	For any given AoD $\theta$ in the considered region $\cal D$, the corresponding array response of the $n$-th antenna at any frequency $f$ is expressed as
	\begin{equation}
		{a_n}(f,\theta) = {e^{j2\pi f {c_{\alpha}c_{\gamma}\cos\theta x_n /c}},f \in [{f_c} - \frac{B}{2},{f_c} + \frac{B}{2}]}.
	\end{equation}
	Thus, the array response vector of the ULA can be written as
	\begin{equation}
    \mv{a}\left( f,\theta \right) = [{a_1}(f,\theta),...,{a_N}(f,\theta)]^H.
	\end{equation}\par 
    Let $\mv{\omega}$ denote the analog beamforming vector of the transmitter, which is subjected to a constant magnitude on each of its elements and given by
	\begin{equation}
	     {\mv{\omega }} = \frac{1}{{\sqrt N }}{[{e^{j{\vartheta _1}}}, e^{j\vartheta _2}},...,e^{j{\vartheta _N}}]^T,
	\end{equation}
    where $\vartheta _n$ denotes the phase shift of the $n$-th antenna.
	Consequently, the beam gain at the AoD $\theta$ and the frequency $f$ can be derived as
    \begin{equation}
        \begin{aligned}
        G(\mv{\omega}, \mv{r}) &= {| \frac{1}{{\sqrt N }}\sum\limits_{n = 1}^N {e^{j(-\vartheta _n+2\pi   f   {cos\theta c_{\alpha}c_{\gamma}x_n/c)}}| }^2} \\
         &= {\left| {{{\mv{\omega }}^H} \mv{a}\left( {f,\theta} \right) }\right|}^2.
         \end{aligned}
    \end{equation}\par
	Note that in the conventional narrowband transmission, the analog beamforming can be designed to align with the AoD at the center frequency, i.e., $\vartheta_n=2\pi f_c \cos\theta c_\alpha c_\gamma x_n/c$. However, as the bandwidth (i.e., $B$) becomes sufficiently large, the narrowband beamforming design may result in significant beam gain loss and even nulling at specific frequencies. It is also interesting to note that if the ULA is rotated such that $c_{\alpha}c_{\gamma} \cos\theta=0$,
    i.e., the AoD becomes equal to $\pi/2$ after the rotation, the beam gain in (11) becomes regardless of the frequency. In this case, a uniform beam gain of $N$ can be achieved over the entire frequency band by setting $\vartheta_n=0, 1 \le n \le N$. However, this may not apply to other AoDs in $\cal D$. To cater to all AoDs in $\cal D$ with the ULA rotation, we define the following spatial wideband beam gain, i.e.,
    \begin{equation}
        {G_{\min }}(\mv{\omega} ,\mv{r}) = \mathop {\min }\limits_{\theta  \in [{\theta _{\min }},{\theta _{\max }}]} \mathop {\min }\limits_{f \in [{f_c} - \frac{B}{2},{f_c} + \frac{B}{2}]} G(\mv{\omega},\mv{r}),
    \end{equation}
    which captures the worst-case beam gain over all frequencies within $[f_c-B/2,f_c+B/2]$ and all AoDs within $[\theta_{\min}, \theta_{\max}]$. It is worth noting that compared to the wideband beam gain defined in our previous works \cite{ning1} and \cite{wang}, the spatial wideband beam gain in (12) further involves the angular domain, which renders its analysis and optimization more challenging.\par
    Based on the above, we aim to jointly optimize the analog transmit beamforming $\mv{\omega}$ and the rotational angle vector $\mv{r}$ to maximize the spatial wideband beam gain in (12), i.e.,
	\begin{subequations}
		\label{P1}
		\begin{align}
			\text{(P1)}:&\mathop {\max }\limits_{\mv{\omega} ,\mv{r}} {G_{\min }}(\mv{\omega} ,\mv{r}) \\
			\text{s.t.}&\left| {\omega_i} \right| = \frac{1}{{\sqrt N }},{\rm{  }}i = 1,2,...,N,\\
            &\alpha, \gamma \in [0,2\pi],
		\end{align}
	\end{subequations}
    where $\omega_i$ denotes the $i$-th entry of $\mv\omega$. However, (P1) is generally difficult to be optimally solved due to the continuous spatial region and frequency range, the intricate coupling of the $\mv{\omega}$ and $\mv{r}$ in (13a), as well as the unit-modulus constraints. Next, we will propose an AO algorithm to solve it.
	
	\section{Proposed Solution to (P1)}
	In this section, we propose an AO algorithm to solve (P1) by decomposing it into two sub-problems for optimizing $\mv{\omega}$ and $\mv{r}$, respectively, and solve each subproblem via the SCA. To this end, we first reformulate (P1) into a simpler form by merging the spatial and frequency domains into a composite domain.
	
	\subsection{Problem Reformulation}
		 First, we define a spatial-frequency composite variable $\Omega=f\cos\theta$,  $\theta  \in [{\theta _{\min }},{\theta _{\max }}]$ and $f \in [{f_c} - \frac{B}{2},{f_c} + \frac{B}{2}]$.
		Then, it can be verified that $\Omega$ must be located within an interval $[{\Omega} _ - ,{\Omega} _ +] $, where the lower and upper bounds are given as follows:\\
		1) If $\theta _{\min }$,$\theta _{\max }  < 0$, we have
\begin{equation}
    \left\{ {\begin{aligned}
{\overline \Omega _ - } = \frac{2\pi d}{c} \cos \left( {{\theta _{\min }}} \right)   ({f_c} - \frac{B}{2}),\\
 {\overline \Omega _ + } = \frac{2\pi d}{c} \cos \left( {{\theta _{\max }}} \right)   ({f_c} + \frac{B}{2});
\end{aligned}} \right.
\end{equation}
       2) If $\theta _{\min }  <  0$,$\theta _{\max }  \ge 0$ and $\left| {{\theta _{\min }}} \right| < \left| {{\theta _{\max }}} \right|$, we have
		 \begin{equation}
    \left\{ {\begin{aligned}
{\overline \Omega _ - } = \frac{2\pi d}{c} \cos \left( {{\theta _{\min }}} \right)   ({f_c} - \frac{B}{2}),\\
 {\overline \Omega _ + } = \frac{2\pi d}{c} \cos \left( {{\theta _{\max }}} \right)   ({f_c} + \frac{B}{2});
\end{aligned}} \right.
\end{equation}
		    3) If $\theta _{\min } < 0$,$\theta _{\max }  \ge 0$ and $\left| {{\theta _{\min }}} \right| \ge \left| {{\theta _{\max }}} \right|$, we have
		 \begin{equation}
    \left\{ {\begin{aligned}
{\overline \Omega _ - } = \frac{2\pi d}{c} \cos \left( {{\theta _{\max }}} \right)   ({f_c} - \frac{B}{2}),\\
 {\overline \Omega _ + } = \frac{2\pi d}{c} \cos \left( {{\theta _{\min }}} \right)   ({f_c} + \frac{B}{2});
\end{aligned}} \right.
\end{equation}
		    4) If $\theta _{\min } ,\theta _{\max }  \ge 0$, we have
		 \begin{equation}
    \left\{ {\begin{aligned}
{\overline \Omega _ - } = \frac{2\pi d}{c} \cos \left( {{\theta _{\max }}} \right)   ({f_c} - \frac{B}{2}),\\
 {\overline \Omega _ + } = \frac{2\pi d}{c} \cos \left( {{\theta _{\min }}} \right)   ({f_c} + \frac{B}{2}).
\end{aligned}} \right.
\end{equation}
	Accordingly, the array response vector in (9) is rewritten as
		\begin{equation}
			\mv{a}(\overline{\Omega}, \alpha, \gamma ) = [1,{e^{j \overline \Omega  {c_\alpha }   {c_\gamma } }},...,{e^{j(n-1)\overline \Omega  {c_\alpha }  {c_\gamma }}}].
		\end{equation}\par
	As a result, (P1) can be reformulated as a wide-beam coverage problem in this composite domain, i.e.,
    \begin{subequations}
		\begin{align}
			\text{(P2)}:&\mathop {\max }\limits_{{\mv{\omega }},\mv{r} } {\rm{ }}\mathop {\min }\limits_{{\overline{\Omega }} \in [{\overline{\Omega} _ - },{\overline{\Omega} _ + }]} {\rm{ }}G(\mv{\omega},\mv{r}),\\
			\text{s.t.}&\left| {\omega_i} \right| = \frac{1}{{\sqrt N }},{\rm{  }}i = 1,2,...,N,\\	
            &\alpha, \gamma \in [0,2\pi].
		\end{align}
	\end{subequations} \par
    Furthermore, it is also noted that the rotation angles only affect the objective function of (P2) in a single term, i.e., $c_{\alpha}c_{\gamma}$. Hence, it suffices to optimize $\mu \triangleq c_{\alpha}c_{\gamma}$ in (P2) as a rotation coefficient, with $-1 \le \mu \le 1$. Thus, the array response in (18) can be further simplified as
        \begin{equation}
            {\mv{a}}(\overline \Omega  ,\mu ) = [1,{e^{j\overline \Omega  \mu }},...,{e^{j(n - 1)\overline \Omega  \mu }}].
        \end{equation}\par
     Moreover, we discretize the continuous spatial-frequency interval $[\overline \Omega_{\min},\overline \Omega_{\max}]$ in (P2) into a set of discrete values. Let $L$ represent the number of the sampling points. Then, the $l$-th sampling point can be expressed as
	\begin{equation}
		\overline \Omega_l = \overline \Omega _{\min } + \frac{{l - 1}}{{L - 1}}({\overline \Omega_{\max}-\overline \Omega_{\min}}),{l} = 1,2,...,{L}.
	\end{equation}\par
    Based on the above, (P2) can be simplified as
    \begin{subequations}
		\begin{align}
			\text{(P3)}:&\mathop {\max }\limits_{\mv{\omega} ,\mu ,\varsigma } \varsigma \\
			\text{s.t. }&G(\overline {\Omega }_l , \mv{\omega} ,\mu ) \ge \varsigma ,\forall l,\\
			&-1 \le \mu  \le 1,\\
			&\left| {\omega_i} \right| = \frac{1}{{\sqrt N }},{\rm{  }}i = 1,2,...,N,
		\end{align}
	\end{subequations} 
    where an auxiliary variable $\varsigma$ is introduced. As compared to (P1), the reformulated problem (P3) is more tractable to solve by merging the frequency and spatial domains into a composite domain and introducing the auxiliary variable $\mu$ and the discretization. However, it is still a non-convex problem due to the coupling of $\mv{\omega}$ and $\mu$. Next, we present the AO algorithm to solve it.

	\subsection{Optimizing the Analog Beamforming with a Given $\mu$}
	In this subsection, we aim to optimize the analog transmit beamforming vector $\mv {\omega}$ with a given $\mu$.
	To deal with the non-convex constraint (22b), we first rewrite the objective its left-hand side as
	\begin{equation}
		\begin{array}{l}
			G(\overline {\Omega }_l ,\mv{\omega}, \mu ) = {{\mv{\omega }}^{H}}{\mv{a(}}\overline {\Omega }_l { ,\mu )}   {\mv{a(}}\overline {\Omega }_l {,\mu }{{ \mv{)}}^{H}}{\mv{\omega }} = \text{Tr}({\mv{V}_l} \mv{W}{\rm{)}},
		\end{array}
	\end{equation}
	where ${\mv{V}_l} = {\mv{a}}(\overline {\Omega }_l {,\mu )}   {\mv{a(}}\overline {\Omega }_l {,\mu }{{ \mv{)}}^{H}}$ and ${ \mv{W}} = {{\mv{\omega }}^{H}}{\mv{\omega }}$. Then, for any given $\mu$, (P3) can be reformulated as
	\begin{subequations}
		\begin{align}
			\text{(P3.1)}: &\mathop {\max }\limits_{ \mv{W} } \varsigma \\
			\text{s.t. }&\text{Tr}({\mv{V}_l} \mv{W}) \ge \varsigma ,\forall l,\\
			&{ \mv{W}(n,n)} = \frac{1}{ N },n \in N,\\
			&{ \mv{W}} \succeq 0,\\
            &\text{rank}(\mv{W}) = 1.
		\end{align}
	\end{subequations} \par
    To deal with the rank-one constraint rank$(\mv{W})=1$, we note that
	\begin{equation}
		\text{rank}({ \mv{W}}) = 1 \Leftrightarrow f( \mv{W}) \buildrel \Delta \over = {\left\| { \mv{W}} \right\| }_ * - {\left\| { \mv{W}} \right\|}_2 = 0,
	\end{equation}
	where the nuclear norm ${\left\| { \mv{W}} \right\| }_ *$ equals to the summation of all singular value of matrix ${ \mv{W}}$, and the spectral norm ${\left\| { \mv{W}} \right\|_2}$ equals to the largest singular value of ${ \mv{W}}$. Then, the objective function in (P3.1) can be reformulated as
	\begin{equation}
		\mathop {\max }\limits_{ \mv{W},\varsigma} \varsigma  - \rho f( \mv{W}),
	\end{equation}
	where $\rho > 0 $ is the penalty parameter which ensuring the objective function is small enough if ${\left\| { \mv{W}} \right\|_*} - {\left\| { \mv{W}} \right\|_2} \ne 0$. However, the objective function is still non-convex due to ${\left\| { \mv{W}} \right\|_2}$ represents the maximum singular value of $ \mv{W}$. Nonetheless, we could utilize the SCA algorithm to get a locally optimal solution. In particular, for any given local point ${{ \mv{W}}^{(i)}}$ in the $i$-th SCA iteration, we replace $f( \mv{W})$ as its first-order Taylor expansion, i.e.,
	\begin{equation}
	\begin{aligned}
		f({ \mv{W}}) &\ge \widetilde f({ \mv{W}}|{{ \mv{W}}^{(i)}}){\rm{  }} \buildrel \Delta \over = {| { \mv{W}} |_*} - ({|| {{{ \mv{W}}^{(i)}}} ||_2}\\ &+ {\rm{Re}}(\rm{Tr}(({\partial _{{{ \mv{W}}^{(i)}}}}{\left\| { \mv{W}} \right\|_2})({ \mv{W}} - {{ \mv{W}}^{(i)}})))),
	\end{aligned}
	\end{equation}
	 where the sub-gradient of ${\left\| { \mv{W}} \right\|_2}$ can be computed as ${\partial _{{{ \mv{W}}^{(i)}}}}{\left\| { \mv{W}} \right\|_2} = { \mv{s}} {{\mv{s}}^H}$, where $s$ denotes the eigenvector corresponding to the largest eigenvalue of $\mv{W}^{(i)}$ \cite{passive}. Based on the above, the optimization problem of $ \mv{W}$ in the $i$-th SCA iteration is given by
	 	\begin{subequations}
	 	\begin{align}
	 		\text{(P3.2)}: &\mathop {\max }\limits_{{ \mv{W}},{ \varsigma }} \varsigma  - \rho \widetilde f{ \mv{(W}|}{{ \mv{W}}^{(i)}}{\rm{)}}\\
	 		\text{s.t. }&\text{Tr}({ \mv{V}_l \mv{W}}) \ge \varsigma ,\forall l,\\
	 		&{ \mv{W}(n,n)} = \frac{1}{ N },n \in N,\\
	 		&{ \mv{W}} \succeq 0.
	 	\end{align}
	 \end{subequations}
	 It can be seen that the objective function (P3.2) is currently a linear function of $\varsigma$ and $\mv{W}$; thus, (P3.2) can be optimally solved by the interior-point algorithm. Based on our simulation results, by properly setting the value of penalty parameter $\rho$, the converged solution of $ \mv{W}$ is always rank-one. Then, we proceed to the $(i+1)$-th SCA iteration for $ \mv{W}$ by updating ${{ \mv{W}}^{(i + 1)}}$ as the optimal solution to (P3.2). Finally, we perform the singular value decomposition (SVD) on ${ \mv{W}} = { \mv{U}}_1^H{ \mv{\Lambda }}{{ \mv{U}}_2}$, where ${ \mv{U}}_1$, $ \mv{\Lambda }$ and ${{ \mv{U}}_2}$ are the left eigenvector matrix, the diagonal matrix of the singular values and the right eigenvector matrix of $ \mv{W}$, respectively. The analog beamforming solution to (P3.1) can be derived as $\mv{\omega}  = 1/\sqrt N {e^{j\arg ({ \mv{U}}_1^H   \sqrt { \mv{\Lambda }}    \mv{q})}}$ accordingly, where $\mv{q}$ is the left singular vector corresponding to the largest singular value of $ \mv{W}$.
	
	\subsection{Optimizing the Rotation Coefficient with a Given $\mv{\omega}$}
	Next, we optimize the rotation coefficient $\mu$ with a given analog transmit beamforming $\mv{\omega}$. First, to tackle the non-convex constraint (22b), we expand $G(\overline{\Omega}_l,\mv{\omega},\mu)$ into the following form
	\begin{equation}
		\begin{aligned}
		    G(\overline \Omega_l  ,\mu ) &= {\left| \frac{1}{\sqrt{N}}{\sum\limits_{n = 1}^N {e^{-j\omega_n} {e^{j\overline {   \Omega }_l    \mu }}} } \right|^2}\\
			 &= \sum\limits_{m = 1}^N {\sum\limits_{n = 1}^N {\frac{1}{N}\cos (\mu(\overline \Omega_l   (n - m) - \Delta \phi_{n,m}))}},
		\end{aligned}
	\end{equation}
	where $\Delta \phi_{n,m}  = \omega_n-\omega_m$. Then, we construct a convex surrogate function to solve the non-convex problem according to the Taylor expansion\cite{mamulti}. For a given $z_0\in\mathbb{R}$, the second-order Taylor expansion of a cosine function is given by
	\begin{equation}
		\cos (z) \approx \cos ({z_0}) - \sin ({z_0})(z - {z_0}) - \frac{1}{2}\cos ({z_0}){(z - {z_0})^2},
	\end{equation}
	since $\cos ({z_0}) \le 1$ and ${(z - {z_0})^2} \ge 0,$ we can construct a quadratic function $g(z|z_0)$ to replace $\cos(z)$ as 
	\begin{equation}
		\cos (z) \ge \cos ({z_0}) - \sin ({z_0})(z - {z_0}) - \frac{1}{2}{(z - {z_0})^2}.
	\end{equation}
	
	Based on the above, for any local point $\mu^{(i)}$ in the $i$-th SCA iteration for $\mu$, we can construct a surrogate function of $G(\overline {\Omega}_l,\mu)$ as
	\begin{equation}
		G(\overline \Omega_l   ,\mu ) \ge A_l    {\mu ^2} + B_l    \mu  + C_l ,
	\end{equation}
	where $A_l $, $B_l $ and $C_l $ respectively represent square-term, primary-term, and constant coefficients, which are given by
	\begin{subequations}
    \begin{align}
	A_l  =& \sum\limits_{n = 1}^N {\sum\limits_{m = n}^N { - \frac{1}{2}} } {(\overline \Omega_l)^2}   {(n - m)^2},\\
	B_l  =& \sum\limits_{n = 1}^N {\sum\limits_{m = n}^N {{\mu^{(i)}}} } {({ \overline \Omega_l  })^2}   {(n - m)^2} \\ \nonumber
    -& \sin [ \overline \Omega_l  (n - m){\mu^{(i)}} - \Delta\phi_{m,n} ]    \overline \Omega_l  (n - m),\\ 
	C_l =& \sum\limits_{n = 1}^N {\sum\limits_{m = n}^N {\frac{1}{2}} } {({ \overline \Omega_l  })^2}   {(n - m)^2}{\mu^{(i)}}^2\\ \nonumber
	+& \sin [ \overline \Omega_l  (n - m){\mu^{(i)}} - \Delta\phi_{m,n} ]   \overline \Omega_l  (n - m){\mu^{(i)}}\\ \nonumber
    +& \cos [ \overline \Omega_l  (n - m){\mu^{(i)}} - \Delta\phi_{m,n} ].
    \end{align}
	\end{subequations}\par
    It is noted that $A_l$ is always no larger than zero and is independent of $\overline{\Omega}_l$. As such, constraint (33) is a quadratic contraint w.r.t. the rotation coefficient $\mu$. Therefore, in the $i$-th SCA iteration for optimizing $\mu$, the following optimization problem should be solved, i.e.,
    \begin{subequations}
	 	\begin{align}
	 		\text{(P3.3)}: &\mathop {\max }\limits_{\mu,\varsigma} \varsigma  \\
	 		\text{s.t. }&A_l {\mu ^2} + B_l  \mu  + C_l  \ge \varsigma ,\forall l,\\
                &-1 \le \mu \le 1.
	\end{align}
    \end{subequations}\par
    It appears that (P3.3) is a classic  quadratically constrained quadratic program (QCQP) problem, which can be effectively solved via the interior-point algorithm. Then, we can proceed to the $(i+1)$-th SCA iteration for $\mu$ by updating the local point $\mu^{(i)}$ as the optimal solution to (P3.3). Finally, we reconstruct the rotational angles $\alpha$ and $\gamma$ from the optimized $\mu$ based on $\mu=c_{\alpha}c_{\gamma}$, which has an infinite number of feasible solutions. One feasible solution is by setting $\alpha = 0$, and 
    \begin{equation}
        \gamma = \text{arccos}(\mu).
    \end{equation}
	
	\subsection{Overall Algorithm and Convergence Analysis}
	The overall AO algorithm can be executed as follows. Let $\mv{\omega}(j-1)$ and $x(j-1)$ denote the values of $\mv{\omega}$ and $\mu$ at the beginning of the $j$-th AO iteration. Then, in this AO iteration, we first optimize $\mv{W}$ via SCA with fixing $\mu=\mu(j-1)$ and $\mv{W}^{(0)}=\mv{\omega}(j-1)\mv{\omega}^H(j-1)$ and obtain $\mv{\omega}(j)$ by performing SVD on the converged $\mv{W}$. Next, we optimize $\mu$ via SCA with fixing $\mv{\omega}=\mv{\omega}(j)$ and $\mu^{(0)}=\mu(j-1)$ and obtain $\mu(j)$ as the converged solution. The $(j+1)$-th AO iteration follows. The overall procedures of the AO algorithm are summarized in Algorithm 1. \par
    \begin{algorithm}[!h]
	\caption{Proposed AO algorithm to solve (P1)}
	\label{alg:AOA}
	\renewcommand{\algorithmicrequire}{\textbf{Input:}}
	\renewcommand{\algorithmicensure}{\textbf{Output:}}
	\begin{algorithmic}[1]
		\REQUIRE  $\mv{\omega}(0)$ and $\mu(0)$. 
		\STATE Initialization: $j\leftarrow 1$.
		\WHILE{AO convergence is not reached}
		\STATE Initialize $i\leftarrow 0$ and update $\mv{W}^{(0)}=\mv{\omega}(j-1)\mv{\omega}^H(j-1)$, and $\mu=\mu(j-1)$.
		\WHILE{SCA convergence for $\mv{W}$ is not reached}    
		\STATE Obtain $\mv{W}^{(i+1)}$ by solving problem (P3.2).
		\STATE Update $i\leftarrow i+1$.        
		\ENDWHILE
		\STATE Obtain $\mv{\omega}(j)$ based on the SVD on $\mv{W}^{(i)}$.        
		\STATE Initialize $i\leftarrow 0$ and update $\mv{\omega}=\mv{\omega}(j)$, $\mu^{(0)}=\mu(j-1)$.
		\WHILE{SCA convergence for $\mu$ is not reached}
		\STATE Obtain $\mu^{(i+1)}$ by solving problem (P3.3).
		\STATE Update $i\leftarrow i+1$.
		\ENDWHILE
		\STATE Update $\mu(j)=\mu^{(i)}$.
		\STATE $j\leftarrow j+1$.
		\ENDWHILE       
        \STATE $\mu = \mu(j)$, $\mv{\omega} = \mv{\omega}(j)$
        \STATE Obtain $\alpha$ and $\gamma$ based on (35)
		\RETURN $\alpha$, $\gamma$ and $\mv{\omega}$.
	\end{algorithmic}
\end{algorithm}
    Next, we show the covergence of our proposed algorithm. Specifically, regarding the SCA for $\mv{W}$, let $v^{(i)} \triangleq \varsigma - \rho f(\mv{W}^{(i)})$ denote the objective value in the $i$-th SCA iteration for $\mv{W}$. Then, the following inequalities hold, i.e.,
    \begin{equation}
    \begin{aligned}
        v^{(i)} &\triangleq \varsigma - \rho f(\mv{W}^{(i)}) \mathop  = \limits^{(a)} \varsigma - \rho \widetilde f({\mv{W}^{(i)}}|{\mv{W}^{(i)}})\\
        &\mathop  \le \limits^{(b)} \varsigma - \rho \widetilde f({\mv{W}^{(i + 1)}}|{\mv{W}^{(i)}}) \mathop  \le \limits^{(c)} \varsigma - \rho f({\mv{W}^{(i + 1)}}) \triangleq v^{(i+1)},
    \end{aligned}
	\end{equation}
	where the equality ($a$) holds since the first-order Taylor expansion in (27) is tight at $\mv{W}^{(i)}$; the inequality ($b$) holds as $\mv{W}^{(i+1)}$ is the optimal solution to (P3.2) and thus maximizes the function $\varsigma - \rho \widetilde f({\mv{W}^{(i)}}|{\mv{W}^{(i)}})$; the inequality ($c$) holds due to (27). Based on the above, the sequence $\{v^{(i)}\}$ is non-decreasing and thus ensured to converge. Similarly, it can be shown that the SCA for optimizing the rotation coefficient $\mu$ also converges, for which the details are omitted for brevity. It follows that the proposed AO algorithm can converge.\par    
	Finally, we analyze the complexity of the proposed algorithm. It can be shown that the complexity of optimizing the analog transmit beamforming $\mv{\omega}$ and the rotational coefficient $\mu$ are both on the order of $\mathcal{O}(\sqrt L N({N^2} + L))$\cite{outage}.
   \begin{figure}[hbtp]
        \centering
        \subfigure[Without antenna rotation] {\includegraphics[width=0.35\textwidth]{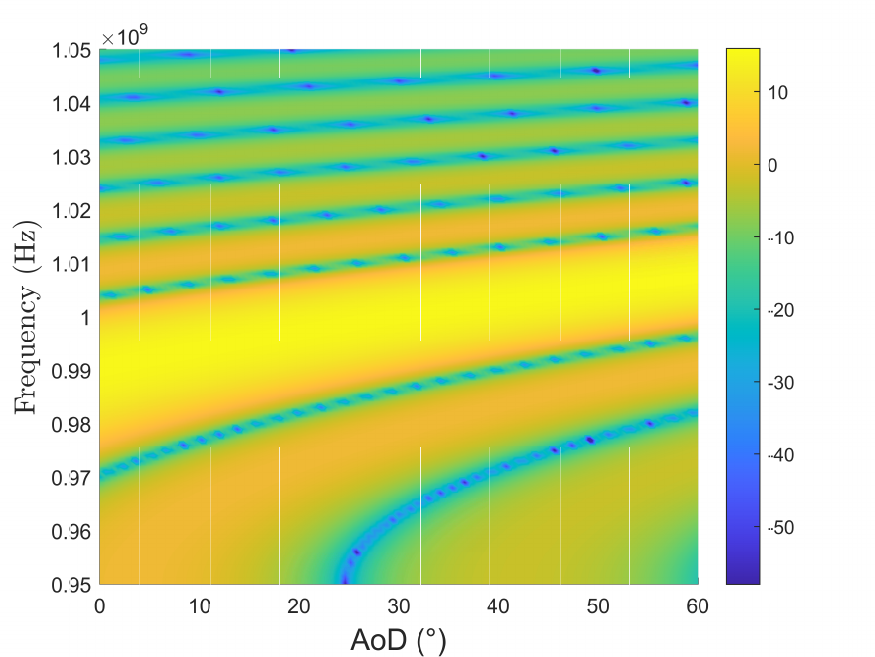}\label{n_6}}
        \quad
        \subfigure[With antenna rotation] {\includegraphics[width=0.35\textwidth]{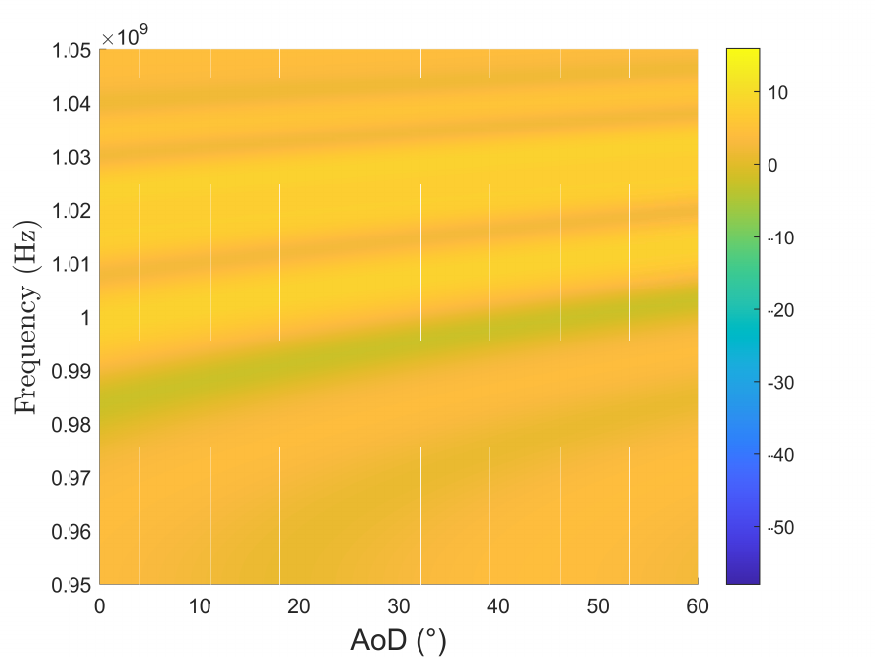}\label{n_8}}
        \caption{Distribution of beam gains over the frequency-spatial domain.}\label{simfig2}\vspace{-12pt}
    \end{figure}

    \subsection{Initialization}
    The converged performance of the AO algorithm depends critically on the initialization. We propose an efficient initialization method to determine $\mu(0)$ and $\mv{\omega}(0)$ in Algorithm 1. For the rotation coefficient, we set it as $\mu(0) = 1$, corresponding to the case without antenna rotation. \par
    To determine $\mv{\omega}(0)$, we consider solving problem (P3.1) directly with $\mu=\mu(0)=1$ via semidefinite relaxation (SDR) by ignoring constraint (24c). Denote by $\mv{W}(0)$ the associated optimal solution to (P3.1), which can be obtained by invoking the interior-point algorithm. Then, we obtain $\mv{\omega}(0)$ by performing Gaussian randomization on $\mv{W}(0)$, i.e.,     
    \begin{equation}
        \mv{\omega}(0)  = \frac{1}{\sqrt{N}} {e^{j\arg ({ \mv{U}}_0^H   \sqrt { \mv{\Lambda}_0 }    \mv{q}_0)}},
    \end{equation}
    where $\mv{U}_0$ and $\mv{\Lambda}_0$ are unitary matrix and diagonal matrix composed by eigenvectors and eigenvalues of the matrix $\mv{W}_0$, respectively, and $\mv{q}_0$ is a random vector following $\mathcal{CN}(0,\mv{I}_N)$. By comparing the objective value of (P3.1) under different realizations of $\mv{q}_0$, we set $\mv{\omega}(0)$ as the one yielding the maximum objective value of (P3.1).\vspace{-4pt}
    
	\section{Numerical Results}
	In this section, we provide numerical results to evaluate the performance of the proposed algorithm with 3D rotatable antenna. The baseband bandwidth is set to $B = 0.1\text{ THz}$ and the center frequency is ${f_c} = 1\text{ THz}$. The number of antannas is $N = 32$, and the inter-antenna spacing is set to half-wavelength. The penalty factor is set as $\rho  = 20$. The range of AoD is from ${\theta _{\min }} = 0$ to ${\theta _{\max }} = 60\text{°}$, which result in ${\overline{\Omega} _ -} = 6.85 \times 10^8$ and ${\overline{\Omega} _ + } = 10.5 \times 10^8.$ \par 
	In Figs. 3(a) and 3(b), we plot the distribution of the beam gain over the frequency-spatial domain without versus with the antenna rotation. In the case without the antenna rotation, the transmit beamforming is aligned to the carrier frequency and the center spatial direction, i.e., ${\mv{\omega}}=\mv{a}(f_c,\frac{\pi}{6})$.  
      It is observed from Fig. 3(a) that there exist several deep-fading regions (as marked in the dark color) without the antenna rotation, even if the analog beamforming design has accounted for the spatial coverage. Particularly, the deep fading becomes more significant when the frequency becomes high, due to the more severe beam squint effects. In contrast, an approximately flat beam gain (as marked in a uniformly yellow color) is observed from Fig. 3(b) with the antenna rotation. In particular, the proposed scheme can raise the minimum beam gain by 15 dB compared to the conventional scheme without antenna rotation. The above phenomena suggest that the antenna rotation can be an effective approach to mitigate ``beam-squint" effects.\par 
      Next, we plot in Fig. 4 the max-min beam gain by different schemes over the frequency-spatial composite domain, i.e., $[\overline \Omega_{-}$,$\overline \Omega_{+}]$. We consider the following three benchmark schemes in addition to our proposed scheme with joint beamforming and antenna rotation optimization.\par 
        \begin{figure}[t]
		\centerline{\includegraphics[width=0.48\textwidth]{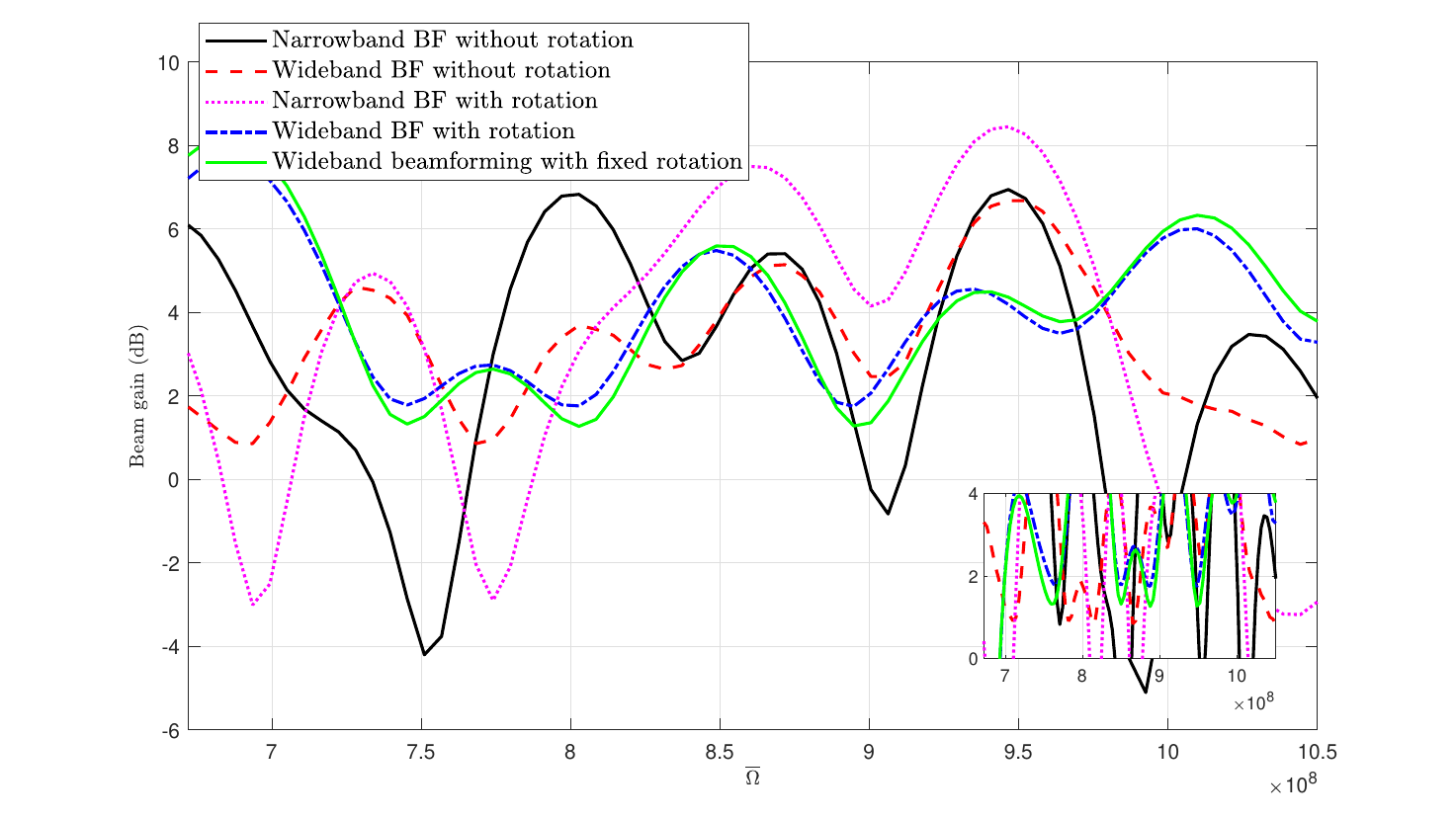}}
		\caption{Optimized beam patterns over the spatial-frequency composite domain by different schemes.}\vspace{-14pt}
		\label{fig3}
	\end{figure} 
	1) {\it Narrowband beamforming without rotation} (Benchmark 1): The rotation coefficient is set as $\mu_0=1$, and the transmit beamforming is obtained by solving (P1) with $\mu=1$ and $B=0$.\par
    2) {\it Wideband beamforming without antenna rotation} (Benchmark 2): The transmit beamforming $\mv{\omega}$ is optimized based on the SCA algorithm as in Section \uppercase\expandafter{\romannumeral3}-A, with $\mu_0=1$.\par 
	3) {\it Narrowband beamforming with antenna rotation} (Benchmark 3): The rotation angle $\mu$ is optimized based on the SCA algorithm as in Section \uppercase\expandafter{\romannumeral3}-B, while the transmit beamforming is set the same as that in Benchmark 1.\par 
    4) {\it Wideband beamforming with fixed antenna rotation} (Benchmark 4): The transmit beamforming $\mv{\omega}$ is optimized based on SCA algorithm as in Section \uppercase\expandafter{\romannumeral3}-A, with the rotation angle fixed as the center of the angular range, i.e., 30$^{\circ}$.\par
    It is observed from Fig. 4 that our proposed AO algorithm can achieve a larger max-min beam gain over the entire composite domain compared to the other four benchmark schemes, i.e., $12$ dB, $9$ dB, $2$ dB and $1$ dB higher than Benchmarks 1, 2, 3 and 4, respectively. This observation implies that incorporating antenna rotation can effectively mitigate the beam squint. Particularly, it is observed under that Benchmark 1, there exist significant fluctuations over the composite domain, although its beamforming design has been optimized to generate a wide beam in the spatial domain. Meanwhile, it is also observed that both Benchmarks 2, 3, and 4 can effectively enhance the max-min beam gain compared to Benchmark 1 by further performing beamforming or antenna rotation optimization. Furthermore, Benchmark 2 is observed to outperform Benchmark 3, which indicates that beamforming optimization may still play a more dominant role than antenna rotation optimization for beam squint mitigation. The performance gap between the proposed scheme and Benchmark 4 indicates that the optimal antenna rotation may not align the antenna boresight with the center AoD, as this cannot optimally balance the power distribution in the angular domain given the correlation among the steering vectors for different AoDs after the antenna rotation.

	\section{Conclusion}
	In this paper, we investigated the application of 3D rotatable antennas to THz beam squint mitigation in wideband wide-beam coverage. We aimed to jointly optimize the analog beamforming and 3D rotational angles of a rotatable antenna array to maximize the minimum beam gain over the desired spatial and frequency ranges. We reformulated this problem into an equivalent wide-beam generation problem in the spatial-frequency composite domain and proposed the AO and SCA algorithms to solve it efficiently. Numerical results demonstrated that our proposed scheme significantly outperforms the conventional FPAs without antenna rotation and conventional narrowband beamforming designed to generate spatially wide beam only.

\bibliographystyle{unsrt}

\end{document}